# Non-Perturbative Renormalisation of Composite Operators

G.Martinelli[a], M.Paciello[a], S.Petrarca[a], C.Pittori[a], C.T.Sachrajda[b], B.Taglienti[a], M.Testa[a] and A.Vladikas[c]

[a] Dip. di Fisica, Univ. di Roma "La Sapienza", Italy
[b] Dept. of Physics, University of Southampton, Southampton SO9 5NH, UK
[c] Dip. di Fisica, Univ. di Roma "Tor Vergata", Italy

Talk presented by C.T.Sachrajda

It is shown that the renormalisation constants of two quark operators can be accurately determined (to a precision of a few percent using 18 gluon configurations) using chiral Ward identities. A method for computing renormalisation constants of generic composite operators without the use of lattice perturbation theory is proposed.

## 1. Introduction

In this talk I will report on attempts to develop techniques for the evaluation of renormalisation constants of composite operators, without the use of lattice perturbation theory. These constants are needed to obtain physical results for quantities such as mesonic decay constants and weak and electromagnetic form-factors from lattice computations of the corresponding matrix elements. Although much of the discussion will be quite general, the specific applications considered below will be to matrix elements of the improved operators [1]

$$\bar{\psi}(x)\left(1+\frac{ra}{2}\gamma\cdot\overleftarrow{D}\right)\Gamma\left(1-\frac{ra}{2}\gamma\cdot\overrightarrow{D}\right)\psi(x) \quad (1)$$

computed using the Sheihkoleslami and Wohlert ($SW$) "improved" fermion action [2]. The discretisation errors in these matrix elements are of $O(\alpha_s a)$ [1] (as compared to $O(a)$ errors in matrix elements computed using the standard Wilson fermion action).

In the next section I will review the use of chiral Ward identities for the determination of renormalisation constants. [3]. The techniques used in sec.2 are generalisations of those proposed in refs. [4] and [5] for simulations with the Wilson fermion action. In section 3 I will briefly describe the status of our attempts to compute the renormalisation constants of general lattice operators. The numerical results presented in this talk were obtained from a simulation in which 18 independent gluon configurations were generated on a $16^3\times 32$ lattice at $\beta=6.0$.

## 2. Chiral Ward Identities

### 2.1. The Conserved Vector Current

In the theory defined by the $SW$ fermion action there is a vector current, $V_\mu^{CI}$, which is both conserved and improved [6]. $V_\mu^{CI}$ is conserved, and so has no perturbative corrections in forward matrix elements, i.e. $Z_{V^{CI}}=1$, and it is improved so that it has no $O(a)$ corrections even in non-forward matrix elements. Defining the "improved" vector current, $V^I$, as that given in eq.(1) with $\Gamma=\gamma^\mu$, its renormalisation constant is given by

$$Z_V = \frac{\langle f|V^{CI}|i\rangle}{\langle f|V^I|i\rangle} \quad (2)$$

$Z_V V^I$ is the correctly normalised vector current [5,7]. In the simulation at $\beta=6.0$, with $|i\rangle=|\rho\rangle$ and $|f\rangle=|0\rangle$, we find $Z_V=0.824(2)$ at $\kappa=0.1425$ (for which the pion has a mass of about 900 MeV) and $Z_V=0.833(1)$ at $\kappa=0.1410$ (for which the pion has a mass of about 1100 MeV). The statistical errors in the ratios (2) tend to be very small. These results can be compared to that obtained from perturbation theory

$Z_V = 1 - 0.10g^2 \simeq 0.83$ [8], where the numerical value was estimated using a boosted coupling $g^2 = 1.7\,(6/\beta)$.

## 2.2. The Renormalisation Constant of the Axial-Vector Current

We now determine $Z_V$ and $Z_A$ by requiring that the currents $Z_V V^I$ and $Z_A A^I$ satisfy the continuum chiral Ward identities. These identities, derived from the axial flavour transformations $\delta_A \psi(x) = i\alpha_A^f(x)\frac{\lambda^f}{2}\gamma_5\psi(x)$ and $\delta_A \bar\psi(x) = i\alpha_A^f(x)\bar\psi(x)\frac{\lambda^f}{2}\gamma_5$ imply [3]

$$2\rho \int d^4x \int d^3\vec{y}\,\langle P^{I,f}(x) A_\nu^{I,g}(y) V_\rho^{I,h}(0)\rangle =$$
$$-i\left(\frac{Z_V}{Z_A^2} - \rho r a\right) f^{fgl} \int d^3\vec{y}\,\langle V_\nu^{I,l}(y) V_\rho^{I,h}(0)\rangle$$
$$-i\left(\frac{1}{Z_V} - \rho r a\right) f^{fhl} \int d^3\vec{y}\,\langle A_\nu^{I,g}(y) A_\rho^{I,l}(0)\rangle \quad (3)$$

$P^I$ and $A^I$ are the pseudoscalar density and axial-vector current respectively (defined by (1) with $\Gamma = \gamma^5$ or $\gamma^\mu\gamma^5$) and $2\rho$ is given by

$$2\rho = \frac{\partial_4 \sum_{\vec{y}} < A_4^{I,f}(\vec{y},t_y) P^{\dagger I,f}(\vec{0},0) >}{\sum_{\vec{y}} < P^{I,f}(\vec{y},t_y) P^{\dagger I,f}(\vec{0},0) >} \quad (4)$$

where $t_y \neq 0$, so that the operators are separated (in practice, to avoid the effects of contact terms, we take $t_y$ to be greater than a few lattice spacings). $\lambda^f$ and $f^{fgh}$ are the generators and structure constants of the flavour symmetry group respectively. The terms proportional to $\rho r a$ in eq.(3) come from contact terms arising from the use of the equations of motion in defining the improved operators in (1). By requiring that eq.(3) holds at each $t_y$ the combinations $Z_V/Z_A^2$ and $1/Z_V$ (and hence $Z_V$ and $Z_A$) can be determined. $Z_V$ and $Z_A$ have no discretisation errors of $O(a)$. The result for $Z_V$ must, of course, be consistent with that obtained from the determination using eq.(2).

Since the Ward identity in eq.(3) involves contact terms, it may be thought that large momenta are present which might spoil improvement. The integration over $x$ and $\vec{y}$ on the left hand side of eq.(3) ensures that this is not the case (at least in perturbation theory) [3].

Our best estimate for $Z_A$ at $\beta = 6.0$ and $\kappa = 0.1425$ is 1.09(3). Our preliminary value at $\kappa = 0.1410$ is consistent with this ($Z_A = 1.08(2)$). The result from perturbation theory is $Z_A = 1 - 0.02g^2 \simeq 0.97$. The values of $Z_V$ obtained using three-point functions (with $\nu = \rho = 4$) are consistent with those obtained from the two-point functions in subsec. 2.1 but have larger errors ($Z_V = 0.85(3)$ at $\kappa = 0.1425$).

The Ward Identity for the pseudoscalar and scalar densities gives the ratio of the corresponding renormalisation constants for which we find (at $\kappa = 0.1425$) $Z_P/Z_S = 0.60(2)$ (the result from perturbation theory is $(1 - 0.23g^2)/(1 - 0.06g^2) \simeq 0.68$).

## 2.3. $Z_A$ Obtained Using Quark Green Functions

We have also evaluated $Z_A$ by studying the following Ward identity for quark Green Functions:

$$2\rho \int d^4x \int d^3y \langle u_\alpha(y)\,(u(x)\gamma_5 d(x))\,\bar d_\beta(0)\rangle =$$
$$\left(\frac{1}{Z_A} - \rho r a\right) \int d^3y \langle \gamma_5 d(y)\bar d(0) + u(y)\bar u(0)\gamma^5\rangle_{\alpha\beta}$$

where $u$ and $d$ are the up and down quark fields respectively. We have evaluated $Z_A$ using eq.(5) in the Landau gauge and found $Z_A = 1.14(8)$ at $\kappa = 0.1425$ ($Z_A = 1.08(6)$ at $\kappa = 0.1410$). These results are in excellent agreement with those obtained using the Ward identity (3). It is pleasing to obtain results using quark Green's functions with reasonably small errors, and the rôle of Gribov copies in such calculations is being investigated.

## 3. Towards Non-Perturbative Renormalisation of General Composite Operators

The technology of higher order perturbative calculations in continuum QCD is well advanced, two and three loop calculations are now routinely performed. In lattice perturbation theory the calculations are considerably more complicated, and the one-loop coefficients are frequently large. Lepage and Mackenzie have provided considerable insights into the reasons for this [9], and have suggested a procedure for an efficient reorganisation of the perturbation series, based on mean field theory and the partial summation of tadpole graphs. Nevertheless, the ignorance of higher order perturbative coefficients in lattice perturbation theory for renormalisation constants of composite operators represents an important source of uncertainty in lattice computations. I conclude this talk by briefly explaining an exploratory program we are pursuing in an attempt to avoid the need for lattice perturbation theory for matrix elements of general operators[10].

For the purposes of illustration I consider a multiplicatively renormalisable operator $O$. We propose to define the renormalised operator $O_R$ by fixing a renormalisation condition which can be imposed non-perturbatively on the lattice. For example it may be convenient to require that an S-matrix element of $O_R$, taken between quark states at some fixed momentum, takes a specific value. In our first attempts to implement these ideas we evaluate ratios of the form

$$\frac{S^{-1}(p)G_{O_B}(p)S^{-1}(p)}{S^{-1}(p)G_{V_B}(p)S^{-1}(p)} \qquad (5)$$

where $S(P)$ is the quark propagator,

$$G_{O_B}(p) \equiv \int d^4x d^4y \exp(ip \cdot x) \langle \psi(x) O_B(y) \bar{\psi}(0) \rangle$$

$O_B$ is the bare operator (with the lattice cut-off) and $V_B$ is the time component of the bare vector current. The normalisation by the Green function with the vector current is a convenient method of implementing the necessary wave function renormalisation (i.e. multiplying by $\sqrt{Z_\psi}$ for each external line). Our calculations are performed in the Landau gauge. The renormalised operator $O_R$ is defined by imposing the condition that the ratio (5), with $G_{O_R}$ in the numerator, takes a specific value at momentum $p$, (in practice we trace the numerators and denominators of (5) with the corresponding $\gamma$ matrix).To obtain the matrix elements in some other standard continuum scheme, e.g. the $\overline{MS}$ scheme, then requires a calculation in continuum perturbation theory only. Our initial results have encouragingly small errors, here I only present a few sample preliminary results for finite quantities, from our simulation on the $16^3 \times 32$ lattice at $\beta = 6.0$, with $p = (\pi/8, \vec{0})$ and $\kappa = 0.1425$. For the ratios of (5) for the pseudoscalar and scalar densities we find 3.3(4), for the axial current 0.62(5), and for $Z_\psi S^{-1} G_{V_B} S^{-1}$ we find 1.26(7), where for $Z_\psi$ we take the coefficient of $p_4 \gamma^4$ in the quark propagator. We are currently investigating the corresponding results in both lattice and continuum perturbation theory to gain some insights into our results, and their behaviour with momenta in particular [10].

An important question in our program is whether it is possible to make the momenta sufficiently small so that lattice artefacts are negligible ($pa \ll 1$), and yet sufficiently large that the perturbative calculations are valid ($p/\Lambda_{QCD} \gg 1$). This is under investigation.

Although the discussion presented here has been for quark bilinear operators, it is also likely to be very useful for four-quark operators relevant for weak decays, and for operators in the heavy quark effective theory.

## Acknowledgements

GM acknowledges the partial suppROT of the MURST, Italy and INFN and CTS acknowledges support from the Science and Engineering Council UK, through the award of a Senior Fellowship.